\documentclass[journal]{IEEEtran}   
\usepackage{hyperref}
\usepackage{cite}

\usepackage{amssymb}
\usepackage{amsmath}
\usepackage{cleveref}
\usepackage{graphicx}
\usepackage{blindtext}
\usepackage{multicol}
\usepackage{multirow}
\usepackage{caption}
\usepackage[table]{xcolor} 
\definecolor{softgreen}{HTML}{c8e7ff} 
\usepackage{booktabs}

\IEEEoverridecommandlockouts                              

\renewcommand{\baselinestretch}{0.92} 
\usepackage{titlesec}
\titlespacing{\section}{0pt}{6pt}{4pt} 
\titlespacing{\subsection}{0pt}{4pt}{3pt}

\title{\LARGE \bf
HessianForge: Scalable LiDAR reconstruction with Physics-Informed Neural Representation and Smoothness Energy Constraints
}

\author{Hrishikesh Viswanath$^{1}$, Md Ashiqur Rahman$^{1}$, Chi Lin$^{1}$, Damon Conover$^{2}$, Aniket Bera$^{1}$ \\
$^1$ Dept of CS, Purdue University, West Lafayette, IN, USA \\
$^2$ DEVCOM Army Research Laboratory, USA\\
}

\date{\includegraphics[height=2in]{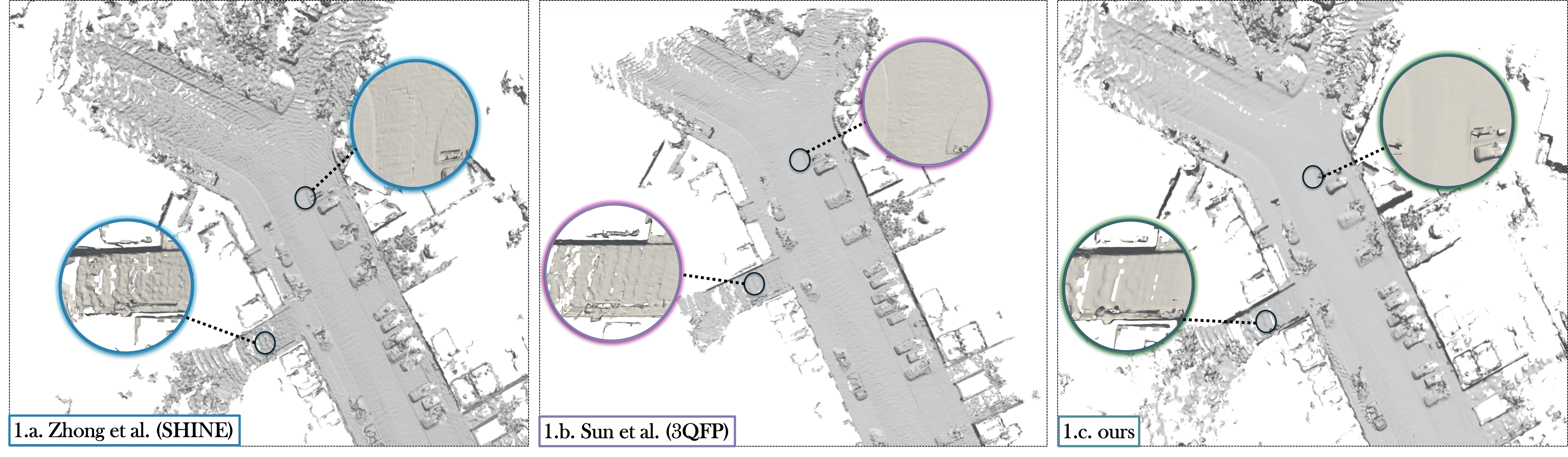}}

\begin{document}
\twocolumn[{%
\renewcommand\twocolumn[1][]{#1}%
\maketitle
\begin{center}
    \centering    
\captionsetup{type=figure}
\includegraphics[width=1\linewidth]{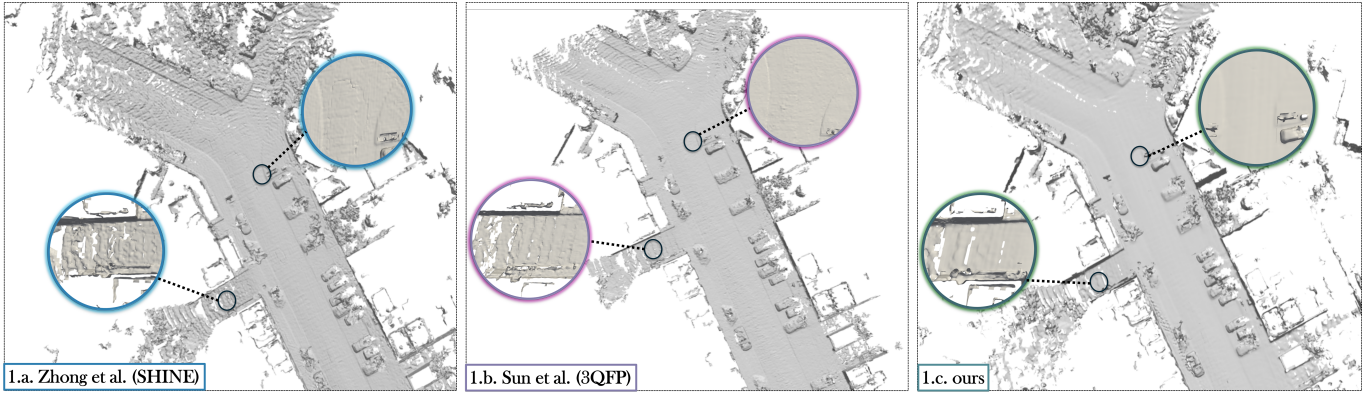}
    \captionof{figure}{\textbf{We present a physics-informed neural implicit representation, optimized on $L_2$-Hessian Energy, for surface mapping from LiDAR point-clouds.} The above figures represent a birds-eye view of \textit{KITTI sequence 00}. Our approach achieves improved surface smoothness and accuracy compared to state-of-the art approaches (1.c.).}
    \label{fig:Introfig}
\end{center}%
}]
\thispagestyle{empty}
\pagestyle{empty}

\begin{abstract}
Accurate and efficient 3D mapping of large-scale outdoor environments from LiDAR measurements is a fundamental challenge in robotics, particularly towards ensuring smooth and artifact-free surface reconstructions. Although the state-of-the-art methods focus on memory-efficient neural representations for high-fidelity surface generation, they often fail to produce artifact-free manifolds, with artifacts arising due to noisy and sparse inputs. To address this issue, we frame surface mapping as a physics-informed energy optimization problem, enforcing surface smoothness by optimizing an energy functional that penalizes sharp surface ridges. Specifically, we propose a deep learning based approach that learns the signed distance field (SDF) of the surface manifold from raw LiDAR point clouds using a physics-informed loss function that optimizes the $L_2$-Hessian energy of the surface. Our learning framework includes a hierarchical octree based input feature encoding and a multi-scale neural network to iteratively refine the signed distance field at different scales of resolution. Lastly, we introduce a test-time refinement strategy to correct topological inconsistencies and edge distortions that can arise in the generated mesh. We propose a \texttt{CUDA}-accelerated least-squares optimization that locally adjusts vertex positions to enforce feature-preserving smoothing. We evaluate our approach on large-scale outdoor datasets and demonstrate that our approach outperforms current state-of-the-art methods in terms of improved accuracy and smoothness. Our code is available at \href{https://github.com/HrishikeshVish/HessianForge/}{https://github.com/HrishikeshVish/HessianForge/}\end{abstract}

\section{INTRODUCTION}
Scene mapping involves constructing a geometric representation of the environment, enabling robots to perceive and interact with their surroundings. It forms a key part of Simultaneous Localization and Mapping (SLAM) problems, which are crucial for autonomous navigation, exploration, and environment monitoring \cite{aulinas2008slam, macario2022comprehensive, andronie2022remote, lluvia2021active, senthilkumaran2023artemis}. These systems rely on various sensors, including LiDAR, cameras, and inertial measurement units (IMUs), to record the observations, which are used to incrementally map the scene. However, due to hardware limitations or limited sensor coverage, the captured point clouds can be sparse and incomplete, particularly in areas where surfaces are only partially observed. This makes it difficult to generate smooth, watertight surface manifolds that accurately represent the scene. Ensuring that the reconstructed surface remains artifact-free and topologically coherent while preserving fine details is a significant challenge in large-scale 3D mapping.

Recent advances in 3D reconstruction leverage neural architectures to map the scene from LiDAR measurements using learnable feature representations. In typical SLAM tasks, the measurements are recorded incrementally as the robot navigates the scene. For memory efficient and scalable feature learning, these methods propose sparse hierarchical tree representations to efficiently encode and query spatial information recorded by the sensors \cite{zhong2023shine, sun20243qfp}. These features are then used by a neural framework to learn signed distance fields (SDF) by incorporating losses derived from partial differential equations (PDE) such as Poisson \cite{vizzo2021poisson, zhong2023shine} or Eikonal equations \cite{zhong2023shine}. Some approaches also leverage kernel-based learning on hierarchical representations, where the surface is represented as the zero-level set of a neural kernel field (NKF), learned as the weighted sum of basis functions defined over a kernel \cite{huang2023neural}. Data-driven approaches, such as NPSR \cite{andrade2023neural}, further generalize these methods by employing neural operators to implicitly learn kernel representations, directly mapping from data without predefined basis functions. Despite these advances, little work has been done in explicitly enforcing geometric smoothness properties in neural representations, to achieve smooth, watertight, and feature-preserving reconstructions.

Smoothness energy functions, such as Laplacian and Hessian energies, are commonly used in graphics and geometry processing for tasks like interpolation, hole-filling, and denoising \cite{rowe2024sharpening, stein2020smoothness}. These energies are defined by the norm of differential operators (i.e. $L_1, L_2$), applied to the surface, where the $L_2$-Laplacian energy minimizes high-frequency variations and the $L_2$-Hessian energy promotes higher-order smoothness \cite{rowe2024sharpening}. Optimizing such $L_2$-energies has been shown to yield smoother geometries. Building on this, we formulate the problem of $3D$ mapping as a deep-learning based energy optimization task constrained by the smoothness energy defined over the surface manifold $\Omega$. Specifically, we propose a novel physics-informed loss function that optimizes the biharmonic partial differential equation (PDE) representation of the $L_2$-Hessian energy.
 
Our proposed neural network framework is inspired by multigrid methods, which are used to solve multiscale PDE problems by propagating information across different resolutions of the domain. These methods facilitate efficient computation by enabling interactions between coarse and fine scales. Motivated by this approach, we introduce a multiscale multilayer perceptron (MLP) architecture for aggregating information across scales. To represent the input, we leverage the octree structure \cite{zhong2023shine}, which inherently partitions 3D space hierarchically, thus providing a natural multigrid discretization of the surface domain $\Omega$. Each level $(l)$ of the octree corresponds to a specific resolution. Unlike multi-pole graph kernel networks, which rely on radius-based connectivity and incur high computational costs in dense graphs, our framework only uses MLPs to aggregate information by utilizing octree node corners as a fixed 8-neighborhood graph, ensuring a per-point computational complexity of $O(l)$, where $l$ is the number of octree levels.

While $L_2$-Hessian energy improves global smoothness, we propose additional fine-grained refinement of local surface details through test-time refinement. Our strategy involves improving the mesh quality by optimizing the mean curvature while preserving vertex connectivity, subject to Laplacian constraints defined on the vertices. The Laplacian operator, defined as the discretized finite difference between a vertex and its neighbors, is used to enforce local smoothness. To enforce these smoothness constraints, we employ a least-squares optimization approach. This approach mitigates the need for model fine-tuning over specific regions, which may otherwise lead to global irregularities due to overfitting. 

Our key contributions can be summarized as follows:
\begin{enumerate}

    \item \textbf{3D mapping as a physics-based energy optimization problem}: We propose a physics-informed approach for modeling large-scale outdoor environments constrained by the biharmonic equation representing the $L_2$-Hessian energy of the surface manifold. 

    \item \textbf{Multi-scale neural representation}: Inspired by multigrid PDE methods, we propose a multi-scale MLP architecture that leverages hierarchical octree representations to refine features across resolutions. 

    \item \textbf{Test-time refinement}: We devise a least-squares optimization strategy for enforcing piece-wise smoothness and improving the triangle quality of the generated mesh.
\end{enumerate}


\section{BACKGROUND}
In this section, we provide an overview of the meshing techniques used in SLAM methods. 

\textbf{LiDAR odometry and meshing:} Conventional LiDAR-based SLAM meshing relies on various techniques for reconstructing surfaces from point-cloud data. These involve correspondence point optimization \cite{zhang2014loam, vizzo2023kiss}, incremental triangulation \cite{lin2023immesh}, Gaussian Process based meshing \cite{jia2024cad, ruan2023slamesh}, surfel based meshing \cite{behley2018efficient}, coverage maps \cite{stachniss2003mapping}, pose estimation based approaches \cite{newcombe2011kinectfusion} and volumetric methods \cite{curless1996volumetric}. 

\textbf{Deep learning based approaches:} With the advent of deep learning, implicit neural representations have emerged as effective surrogates for signed distance function estimation in surface reconstruction. Neural Radiance Fields (NeRF) model surfaces through implicit functions \cite{azinovic2022neural}, while newer approaches leverage Gaussian splatting \cite{chen2023neusg, wu2024recent, matsuki2024gaussian} and triplane-based reconstruction \cite{wang2023pet}. For online SLAM meshing, where fast training and optimization are crucial, methods such as SHINE \cite{zhong2023shine}, NERF-LOAM \cite{deng2023nerf}, 3QFP \cite{sun20243qfp}, introduce efficient tree-based representations to accelerate neural implicit reconstruction.

\textbf{Physics informed surface reconstruction: } Physics-based neural techniques integrate physical priors into neural networks to model the underlying physical properties of a surface with structural and geometric constraints. These methods impose physics-driven regularization, such as smoothness \cite{li2023neuralangelo}, energy minimization \cite{huang2022neural}, or governing differential equations such as Eikonal \cite{wang2024neurodin, zhong2023shine} or Poisson constraints \cite{park2023p}, to guide learning. Another class of techniques involves kernel-based approaches, which use gradient-based data-dependent kernels within the deep learning framework \cite{huang2023neural, williams2021neural, williams2022neural}. Fully data-driven approaches extend this by parameterizing the kernels with neural operators. Neural operators learn function mappings and have shown promise in modeling PDEs \cite{kovachki2023neural, azizzadenesheli2024neural, viswanath2023neural}, but have more recently been used to implicitly model physics in graphics \cite{andrade2023neural, rahman2022pacmo, viswanath2022adafnio, viswanath2024reduced} and robotics \cite{peng2024graph, bhaskara2024trajectory} problems. 

In our work, we extend data-driven kernel methods to design an efficient parametrization of a kernel, via multiscale MLPs, that leverages octree-based data representation for fast feature learning. Moreover, we introduce a physics-informed smoothness energy loss for improved surface quality.

\begin{figure*}[ht!]
    \centering
    \includegraphics[width=1\linewidth]{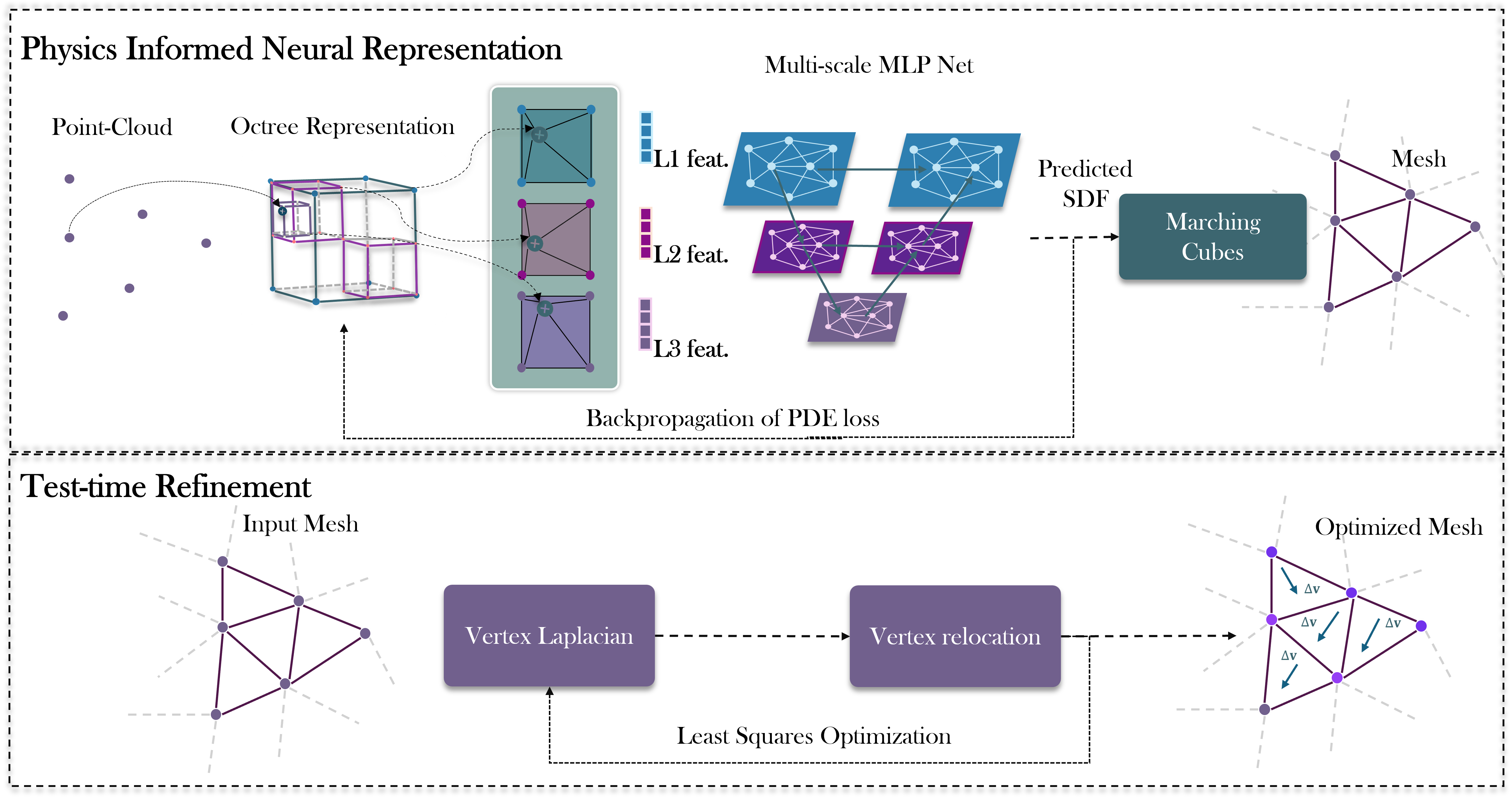}
    \caption{\textbf{Framework architecture:} Split into learnable and non-learnable components, our architecture consists of a neural network architecture trained on $L_2$-Hessian loss and a \texttt{CUDA}-supported test-time refinement module, which includes least squares optimization of vertex Laplacians for fine-grained local smoothing.}
    \label{fig:architecture}
\end{figure*}

\section{PROBLEM SETUP}
In this section, we discuss the problem setting for our mapping framework, which is divided into learnable and non learnable components as shown in \cref{fig:architecture}. The learnable component optimizes a neural network, constrained by smoothness constraints, to represent the signed distance field, while the non-learnable component employs test time refinement for fine-grained local smoothing of the generated mesh. 

Let $\chi = \{x_i \in \mathbb{R}^3, i=1,...,N\}$ be the point cloud sampled from the scene $\Omega$, where, $\Omega$ forms the domain of the optimization problem. We seek a surface mesh $\mathbf{M} = (\mathbf{V}, \mathbf{F})$, with $\mathbf{V}$ vertices and $\mathbf{F}$ faces from the input point cloud $\chi$, such that $\mathbf{M}$ is the surface of the scene determined by the zero-level set of the signed distance field $u$ defined over $\Omega$. Let $u:\mathbb{R}^3\rightarrow\mathbb{R}$ be the signed distance field, whose zero-level set represents the surface, denoted as $\Omega: \{x \in \mathbb{R}^3 \mid u(x) = 0\}$. The surface mesh is obtained by marching cubes on the signed distance field $u$. We aim to learn the mapping $\mathcal{G}_\theta: \chi \rightarrow u$, where the neural network $\mathcal{G}$ is parameterized by the learnable parameter $\theta$. 

\subsection{Smoothness energy as PDE loss}
Smoothness energy is used in graphics and geometry processing problems to enforce smoothness properties over manifolds. This is represented as an energy functional based on differential operators that measure local variation. A key differential measure of smoothness is the \textit{Hessian}, which captures second-order variations in a scalar field. 

For the given signed distance field $u$ defined on the domain $\Omega$, the Hessian of $u$, denoted by $\mathcal{H}_u$, is the second-order differential operator that captures the local curvature of the function by computing the the covariant derivative of the gradient. This is given by $\mathcal{H}_u = \nabla \nabla^\top  u$.

 
The $L_p$-norm of the scalar field defined over a surface $\Omega$ is given by 
$\|u\|_p = (\int_\Omega \|u(x)\|^pdx)^{1/p}$.

The $L_2$-norm, in particular is shown to favor the overall smoothest possible solution \cite{rowe2024sharpening}. Applying this norm to the Hessian defines the $L_2$-Hessian energy, denoted as,
\begin{equation}
    E_{\mathcal{H}} = (\int_\Omega \|\mathcal{H}_u\|^2 dx)^{1/2}
\end{equation}
The square of the Hessian energy functional $E_{\mathcal{H}^2}$ can be interpreted as a strong smoothness regularizer and leads to the formulation of a biharmonic problem or a fourth-order differential equation. To see this connection, we apply the principle of variational calculus to rewrite the minimization problem in terms of the variational formulation \cite{rowe2024sharpening}. This is done by perturbing $u$ with an arbitrary test function $\nu$ as follows,
\begin{equation}
    E_{\mathcal{H}^2({u+\epsilon\nu})} = \int_{\Omega} \sum_{i, j} (\frac{\partial^2 (u + \epsilon \nu)}{\partial x_i \partial x_j})^2 dx .
\end{equation}
Taking the derivative with respect to $\epsilon$ and setting $\epsilon$ to 0 leads to the biharmonic equation. The final form after applying chain-rule is given by
\begin{equation}
    \frac{d}{d\epsilon}E_{\mathcal{H}^2({u+\epsilon\nu})}|_{\epsilon=0} = \int_\Omega \nu \sum_{i,j} \frac{\partial^4 u}{\partial x_i \partial x_j} dx = 0
\end{equation}
This biharmonic equation can be written as $\Delta^2 u = 0$, where $\Delta$ is the Laplace operator. To incorporate this into our optimization, we enforce the condition $\Delta^2 \mathcal{G}_\theta (x) = 0$. The loss function based on this equation is as follows: 
\begin{equation}
    \mathcal{L}_{\mathcal{H}^2} = min_{\theta} \| \Delta^2 \mathcal{G}_\theta (x)\|^2_2
    \label{eq:hessian_loss}
\end{equation}
In practice, we leverage the autograd functionality of \texttt{PYTORCH} and finite difference approximations (FDM) to implement the gradients. The tradeoff between autograd and FDM are discussed in \cref{sec:autograd_tradeoff}. 
\section{FRAMEWORK ARCHITECTURE}
In this section, we will discuss the architecture details and the training setup.

\subsection{Data representation}
We represent our input space using the learnable octree-based representation similar to SHINE \cite{zhong2023shine}. The octree partitions the three-dimensional space by recursively subdividing it into eight octants at each level of refinement. The structure forms a hierarchy of multiple levels $L$, with the smallest level containing octree nodes of size $W$. The subsequent levels in the hierarchy increase in spatial extent such that the $n$-th level contains octree nodes of size $2^n W$, where $n \in \{0, 1, \ldots, L-1\}$.  At every level, a learnable feature vector of dimension $H$ is assigned to each of the $8$ node corners. For a point cloud with $N$ points, the feature tensor is structured as ($N$, $L$, 8, $H$), where each entry corresponds to the feature vector associated with a corner of an octree node at a given level. Feature retrieval is performed via trilinear interpolation between the query point and the feature vectors of the surrounding octree node corners. Following \cite{zhong2023shine}, the octree is implemented using hash tables indexed via Morton codes, enabling efficient querying. 

\subsection{Multi-scale neural network}

Our proposed framework includes a neural implicit representation of the SDF field, which is trained on PDE-based loss functions (\cref{eq:loss}). Inspired by numerical approaches such as the multigrid method, which is a multiscale approach for solving PDEs, our neural network follows a multi-scale MLP architecture for hierarchical feature aggregation, leveraging the hierarchical partitioning of the octree. 

The network architecture mirrors the V-cycle algorithm used in the multigrid method. The V-cycle consists of two main phases: the downward pass and the upward pass. In the downward pass, the solution is progressively restricted to coarser grids. Once the coarsest level is reached, the process reverses in the upward pass, where the solution is iteratively interpolated back to finer grids. This can be written as 
\[
\begin{aligned}
\check v_{l+1} &= K_{l+1, l}\check{v}_l + K_{l+1, l+1}\hat v_{l+1} \\
\hat{v}_{l} &= K_{l, l-1} \hat{v}_{l-1} + K_{l,l}\check{v}_l
\end{aligned}
\]

where, $\check v_l$ refers to the feature vector at level $l$ during the downward pass and $\hat v_l$ refers to the feature vector at level $l$ during the upward pass. We define the kernel operator $K$ as the kernel function applied to the neighborhood of the query point. Specifically, it is expressed as $K = \sum_{x_i \in \mathcal{N}(x)} \alpha_i \mathcal{K}(x, x_i)$, where $x_i$ is a point in the neighborhood $\mathcal{N}(x)$ of $x$. Each level of the upward and downward passes is modeled by a distinct MLP network, where the network parameterizes the kernel functions at that level. This leads to a total of $2l$ MLP networks for an octree of
$l$ levels. Unlike multipole graph kernel networks \cite{li2020multipole} that rely on radius graphs to determine neighborhoods, we redefine the neighborhood structure based on the octree node corners, enforcing a fixed 8-neighborhood aggregation. This design guarantees a constant-cost neighborhood aggregation per node, resulting in an overall complexity of $O(l)$ per-point aggregation per pass.


\subsection{Training} 
To define the complete training and loss setup for the task, we begin by outlining the steps for generating the ground truth, followed by the selection and formulation of the different loss functions used during training.

\begin{table}[ht!]
\vspace{5pt}
\caption{Quantitative results on 3 benchmark LiDAR datasets - \textit{NewerCollege \cite{ramezani2020newer}},  \textit{KITTI \cite{geiger2013vision}} and \textit{MaiCity \cite{vizzo2021poisson}}. We compute accuracy (acc), completeness (comp), Chamfer-L1 distance (C-L1), Precision (Acc Ratio), Recall (Comp Ratio) and F-scores. We use an error threshold of 10cm for \textit{MaiCity} and \textit{KITTI}, while an error threshold of 20cm for \textit{NewerCollege}. In case of \textit{KITTI}, we use a projection of the point-cloud frames as the ground-truth, while the other two datasets include a precomputed ground-truth point-cloud. The values in bold indicate best results.}
\label{tab:main_res}
\resizebox{8.7cm}{!}{
\begin{tabular}{clcccccc}
\toprule
\multicolumn{1}{l}{\textbf{DATA}} & \textbf{MODEL} & \textbf{\begin{tabular}[c]{@{}c@{}}ACC.\\ {[}cm{]}$\downarrow$\end{tabular}} & \textbf{\begin{tabular}[c]{@{}c@{}}COMP.\\ {[}cm{]}$\downarrow$\end{tabular}} & \textbf{\begin{tabular}[c]{@{}c@{}}C-L1\\ {[}m{]}$\downarrow$\end{tabular}} & \textbf{\begin{tabular}[c]{@{}c@{}}ACC. \\ RATIO\\ {[}\%{]}$\uparrow$\end{tabular}} & \textbf{\begin{tabular}[c]{@{}c@{}}COMP. \\ RATIO\\ {[}\%{]}$\uparrow$\end{tabular}} & \textbf{\begin{tabular}[c]{@{}c@{}}F\\ SCORE\\ {[}\%{]}$\uparrow$\end{tabular}} \\ \midrule
\multirow{7}{*}{\textit{\begin{tabular}[c]{@{}c@{}}New\\ College\end{tabular}}} & PUMA & 7.70 & 15.37 & 0.107 & 89.77 & 80.31 & 84.77 \\
 & \begin{tabular}[c]{@{}l@{}}VDB\\ FUSION\end{tabular} & 7.10 & 12.40 & 0.095 & 92.81 & 88.56 & 90.63 \\
 & NKSR & 20.10 & 33.70 & 0.168 & 54.77 & 44.38 & 49.03 \\
 & \begin{tabular}[c]{@{}l@{}}PIN\\ SLAM\end{tabular} & 17.04 & \textbf{10.30} & 0.137 & 62.92 & \textbf{96.67} & 76.22 \\
 & 3QFP & 7.59 & 12.50 & 0.101 & 92.63 & 91.75 & 92.18 \\
 & SHINE & 8.03 & 11.04 & 0.095 & 90.40 & 92.90 & 91.63 \\
  \rowcolor{softgreen} 
 & \textbf{OURS} & \textbf{6.52} & 10.84 & \textbf{0.087} & \textbf{94.67} & 93.25 & \textbf{93.95} \\ \midrule
\multirow{5}{*}{\textit{Kitti}} & ImMesh & 8.65 & 18.67 & 0.077 & 70.20 & 60.33 & 64.89 \\
 & \begin{tabular}[c]{@{}l@{}}PIN\\ SLAM\end{tabular} & 12.33 & 25.81 & 0.09 & 66.43 & 56.70 & 61.18 \\
 & 3QFP & 7.24 & 4.63 & 0.059 & 70.48 & 92.25 & 79.91 \\
 & SHINE & 7.04 & 4.62 & 0.058 & 72.18 & 92.70 & 81.16 \\
 \rowcolor{softgreen} 
 & \textbf{OURS} & \textbf{6.23} & \textbf{4.39} & \textbf{0.056} & \textbf{75.74} & \textbf{93.96} & \textbf{83.87} \\ \midrule
\multirow{7}{*}{\textit{MaiCity}} & ImMesh & 6.78 & 6.10 & 0.077 & 73.06 & 84.42 & 78.33 \\
 & PUMA & 3.50 & 27.58 & 0.056 & 93.01 & 57.45 & 71.02 \\
 & \begin{tabular}[c]{@{}l@{}}VDB\\ FUSION\end{tabular} & 4.00 & 7.10 & 0.044 & 93.01 & 78.22 & 84.97 \\
 & \begin{tabular}[c]{@{}l@{}}PIN\\ SLAM\end{tabular} & 7.39 & 5.50 & 0.0812 & 61.36 & 88.75 & 72.56 \\
 & 3QFP & 3.85 & 5.20 & 0.04 & 92.37 & 90.76 & 91.55 \\
 & SHINE & 3.80 & \textbf{4.70} & 0.042 & 92.60 & \textbf{92.50} & 92.54 \\
  \rowcolor{softgreen} 
 & \textbf{OURS} & \textbf{3.00} & 4.90 & \textbf{0.0398} & \textbf{94.63} & 92.48 & \textbf{93.54} \\ \bottomrule
\end{tabular}}
\end{table}

\textbf{Ground truth SDF}: We follow the steps outlined in \cite{zhong2023shine, sun20243qfp} to compute the ground-truth SDF from LiDAR range sensors. This is achieved by casting rays from the sensor position and calculating the signed distance between each point in the point cloud and the endpoint of the corresponding ray.

\textbf{Loss setup}: We use a combination of three loss functions in our training setup. We use binary cross-entropy loss \cite{zhong2023shine, sun20243qfp} to learn the occupancy probability. This is determined by projecting the SDF into a probabilistic space [0, 1]. We also utilize Eikonal loss, commonly used in surface reconstruction tasks \cite{park2023p ,zhong2023shine}, which enforces the model’s learned surface to satisfy the Eikonal equation, imposing that the gradient of the SDF has a unit norm at every point on the surface. Our loss setup is denoted by 
\[
    \label{eq:loss}
    \mathcal{L} = \lambda_{1} \mathcal{L}_{bce} + \lambda_2\mathcal{L}_{Eikonal} + \lambda_3 \mathcal{L}_{\mathcal{H}^2}
\]
where, $\lambda_1, \lambda_2, \lambda_3$ are tunable hyperparameters. 
\begin{figure*}[th!]
    \centering
    \includegraphics[width=1\textwidth]{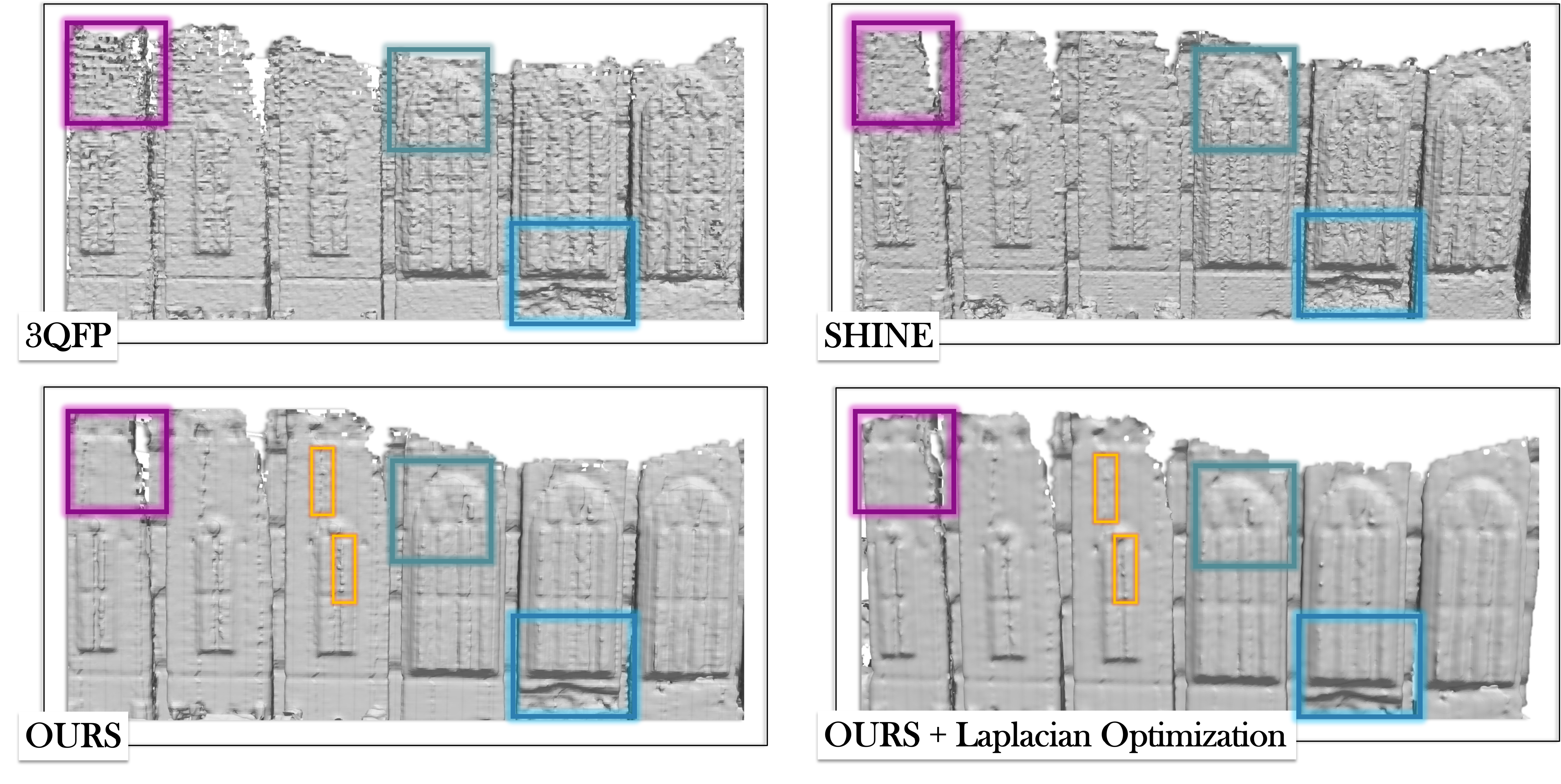}
    \caption{\textbf{Smoothness:} This figure contrasts the quality of reconstruction against baselines on \textit{Newer College}. Our approach achieves higher precision, evidenced by smoother surface.}
    \label{fig:smoothness}
\end{figure*}
\begin{figure}[th!]
    \centering
    \includegraphics[width=8.5cm]{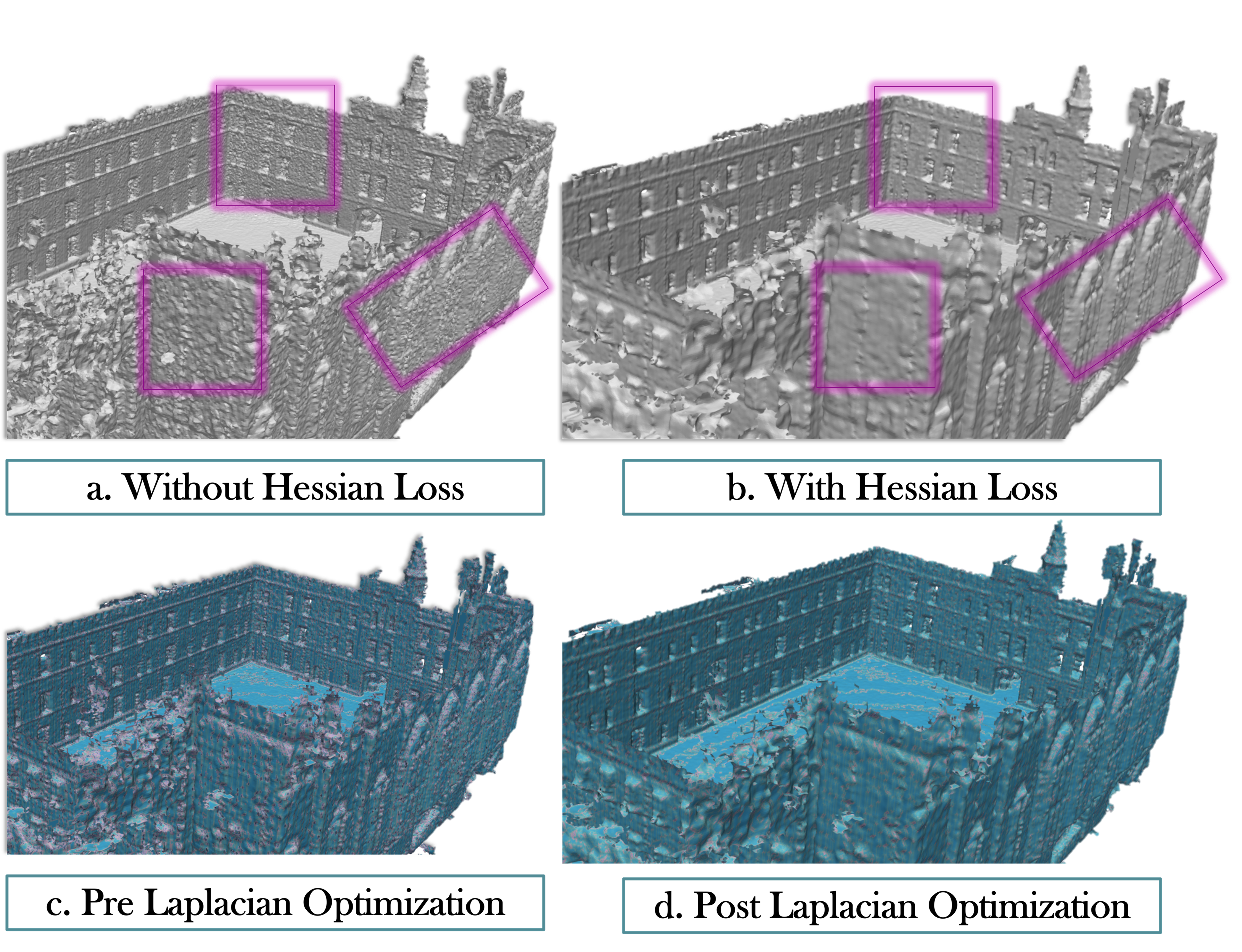}
    \caption{\textbf{Effect of Hessian loss and least squares optimization:} The highlighted areas in (a) and (b) showcase improvements in smoothness due to Hessian energy minimization. (c) and (d) highlight per-vertex Laplacian values. Higher curvature is indicated by gray regions, which is reduced in (d).}
    \label{fig:hessian_opt}
\end{figure}
\begin{figure}[th!]
    \centering
    \includegraphics[width=8.5cm]{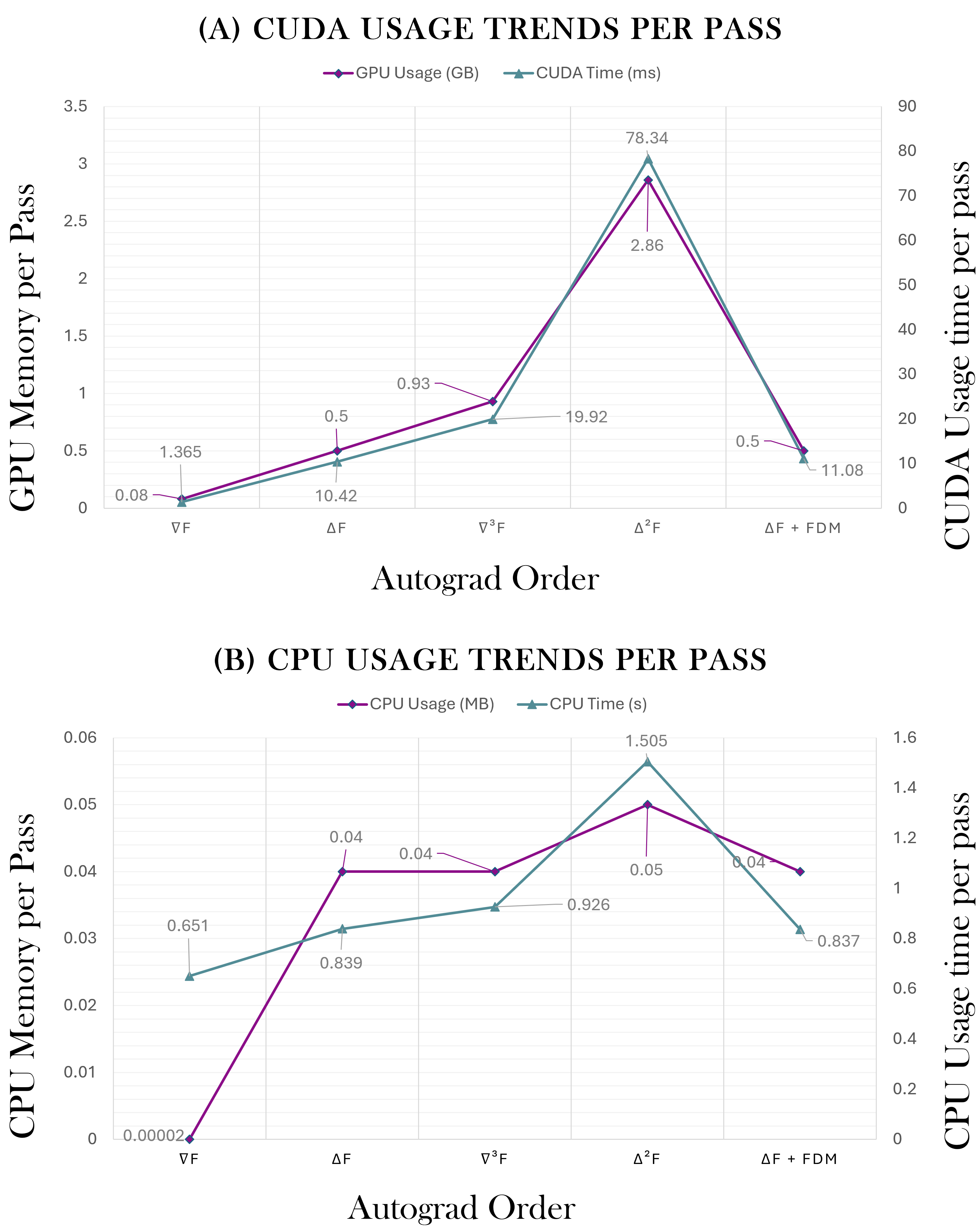}
    \caption{\textbf{Computational overhead of autograd:} The graphs quantify the computational overheads associated with autograd functionality. Combining autograd with FDM approximations reduces the computational overhead. These metrics were computed on RTX 3050 GPU.}
    \label{fig:comp_overhead}
\end{figure}
\section{TEST-TIME REFINEMENT}
While our neural network effectively reconstructs the surface from point-cloud data, the generated output may still contain artifacts due to noise in the data, discretization errors, or the limitations of the learned representation. Fine-tuning the model on local regions could introduce unintended global irregularities. To address this, we introduce a fast differentiable test-time refinement step based on a least-squares optimization strategy to enforce piece-wise smoothness.
\subsection{Laplacians for local smoothing}
To refine the reconstructed mesh, we apply Laplacian smoothing, a technique that redistributes vertex positions to improve local surface smoothness. The Laplacian operator measures the difference between a function's value at a point and its local average, capturing how much a vertex deviates from its neighbors, providing a measure of curvature. In the continuous setting, the Laplacian of a function $f(x)$ is given by $\nabla^2 f$, which represents the second derivative and indicates regions of high curvature. For a discrete mesh representation, we define the discrete Laplacian at a vertex $\mathbf{v_i}$ based on the connectivity of the mesh. Given a set of neighboring vertices $\mathbf{N(i)}$, the discrete Laplacian is computed as $\Delta \mathbf{v_i} = \sum_{j \in \mathbf{N(i)}} w_{ij} (\mathbf{v_i} - \mathbf{v_j})$.\\
Intuitively, this operation moves each vertex toward the centroid of its local neighborhood, reducing noise and ensuring smoother transitions across the surface.

The weights $w_{ij}$ are computed via cotangents, which are commonly used in geometry processing and mesh deformation tasks. The cotangent weight $w_{ij}$ for an edge 
$(i,j)$ is derived from the angles opposite to that edge in the two adjacent triangles. Given two neighboring triangles sharing the edge $(i,j)$, let $\alpha$ and $\beta$
be the angles opposite to $(i,j)$ in each of these triangles. The weight for the edge is computed as $w_{ij} = 1/2(\cot \alpha + \cot \beta)$. 

Given a mesh with $\mathbf{V}$ vertices and $\mathbf{E}$ edges, the Laplacian matrix $\mathbf{L} \in \mathbb{R}^{\mathbf{V}\times\mathbf{V}}$, represents the Laplacian weights for every edge $(i, j)$ connecting a pair of vertices. 
\[
\mathbf{L}_{i,j} = 
\begin{cases}
-\sum_{k \in \mathbf{N}(i)} w_{ik} \quad &\text{if } i=j \\
w_{ij} \quad &\text{if } (i, j) \in \mathbf{E} \\
0 \quad &\text{otherwise}
\end{cases}
\]

The optimization involves minimization of the Laplacian quadratic form $v^\top\mathbf{L}v$, defined as 
$$v^\top\mathbf{L}v = 1/2 \sum_{i, j \in \mathbf{E}_i} (v_i - v_j)^2.$$

\begin{figure}[th!]
    \centering
    \includegraphics[width=8.5cm]{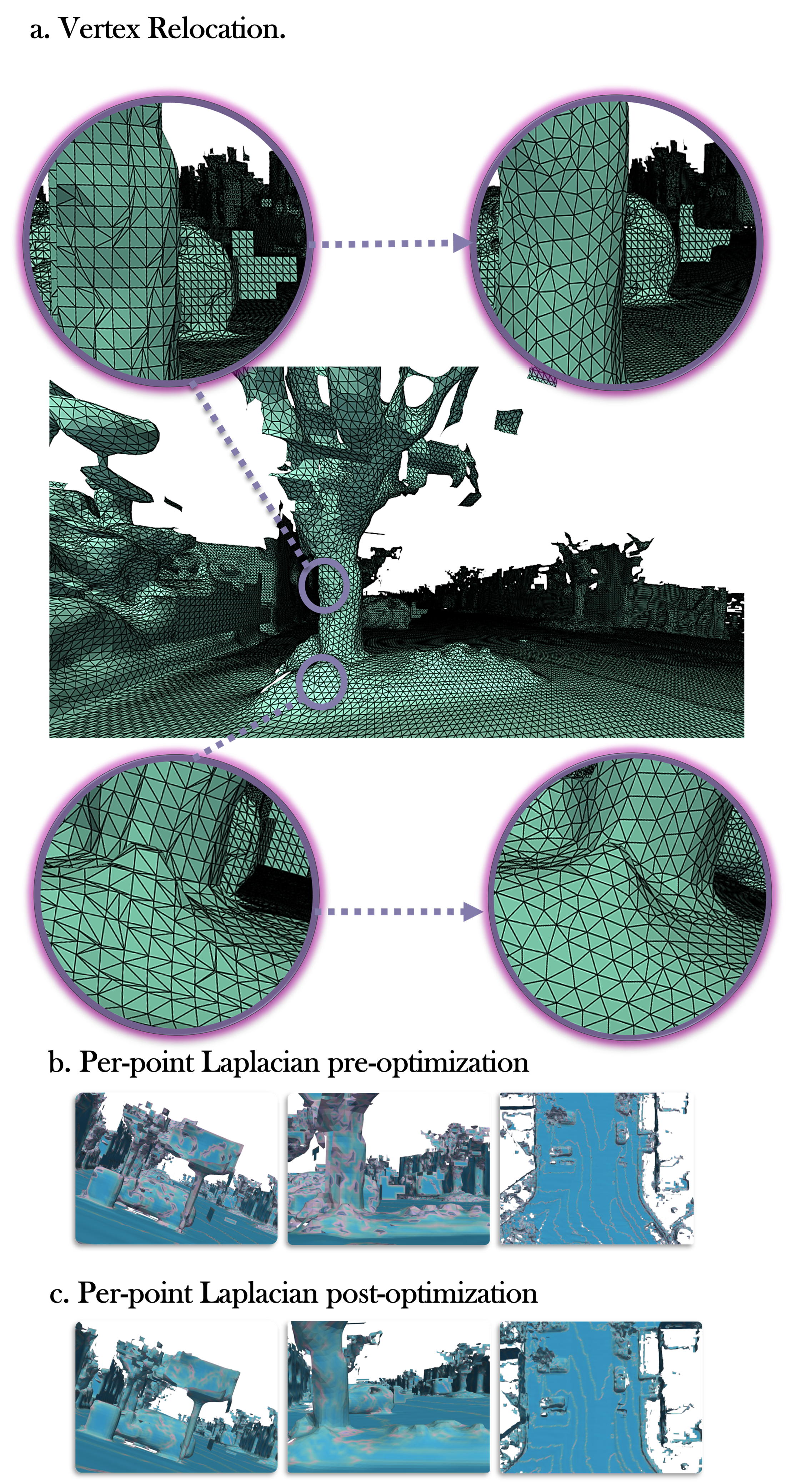}
    \caption{\textbf{Local smoothing via Least squares Optimization:} (a) highlights how vertices are locally optimized. (b) and (c) present per-point Laplacian values before and after optimization. Gray and darker regions indicate higher values of Laplacian or greater curvature.}
    \label{fig:laplace-smoothing}
\end{figure}

This form is used to measure smoothness in computer graphics and spectral analysis \cite{krishnan2013efficient,rahman2025group}. The refinement step can then be expressed as 
\begin{equation}
    v' = v - \eta \frac{\partial v^\top \mathbf{L} v}{\partial v}
\end{equation}
where $\eta$ is a small positive value $0<\eta<1$, denoting the step size. This process displaces the vertex towards its neighbors.
In practice, the displacement is computed by multiplying the Laplacian matrix with the vertices, denoted by $\Delta \mathbf{v} = \mathbf{L}\mathbf{v}$, with the corresponding norm given by $\|\Delta \mathbf{v_i}\| = (\Delta\mathbf{v_{i,x}}^2 + \Delta\mathbf{v_{i,y}}^2 + \Delta\mathbf{v_{i,z}}^2)^{1/2}$ 
This formulation of the Laplacian displacement provides a measure of local curvature variation, which is used to iteratively update vertex positions through least squares optimization:
\begin{equation}
    \mathbf{v_i} = \mathbf{v_i} + \eta \exp(-\|\Delta\mathbf{v_i}\|)(\bar{\mathbf{v}} - \mathbf{v_i})
\end{equation}
where, $\eta$ is the tunable momentum parameter, $\mathbf{\bar{v}}$ is the average location of the neighborhood of the vertex $\mathbf{v_i}$. At each iteration, the Laplacian is recomputed on the updated vertex set, ensuring that vertices shift toward a locally smooth configuration while maintaining global geometric consistency, as seen in \cref{fig:laplace-smoothing}. The exponential weighting prevents excessive displacement in regions of high curvature, preserving sharp edges and geometric details. The entire process is fully differentiable and optimized for \texttt{CUDA}. 

\section{EXPERIMENTS}
\subsection{Dataset}
We evaluate our framework using three outdoor LiDAR datasets: MaiCity, Newer College, and KITTI. The MaiCity dataset is a synthetic dataset generated from a 3D CAD model of an urban environment. It contains 100 frames, which were created by simulating LiDAR scans through ray-casting techniques. The Newer College dataset consists of data captured at New College, Oxford, using a handheld sensor setup, which includes a stereo camera and a 3D LiDAR. This dataset includes 1,500 frames and is accompanied by precise ground truth measurements. Finally, we also benchmark against the KITTI dataset, a widely used dataset for outdoor odometry tasks. KITTI includes sequences captured with a 64-beam LiDAR sensor, along with camera data, and provides near ground truth estimates for incremental mapping in both urban and rural environments.
\subsection{Mapping quality}
We evaluate the quality of the reconstruction using accuracy, completeness, the Chamfer L1 distance, and the F1 score. Accuracy measures the mean distance from the reconstructed surface to the cloud of ground truth points, while completeness quantifies the mean distance from the ground truth points to the reconstructed surface. Chamfer distance measures the bidirectional nearest-neighbor error. We use a 3-level octree with a voxel size of 10 cm. Our MLP architecture is structured with 3 levels, each containing 2 hidden layers with a 128-dimensional feature space. During inference, query points are uniformly sampled and assigned to octree nodes for hierarchical feature aggregation. We present the quantitative results in \cref{tab:main_res}. It can be observed that our approach achieves the highest precision (Acc. Ratio) and F1 scores compared to baseline architectures. We show visual comparisons in \cref{fig:smoothness}.

\subsection{Computational and memory overhead of Hessian energy}
\begin{figure}[th!]
    \centering
    \includegraphics[width=8.8cm]{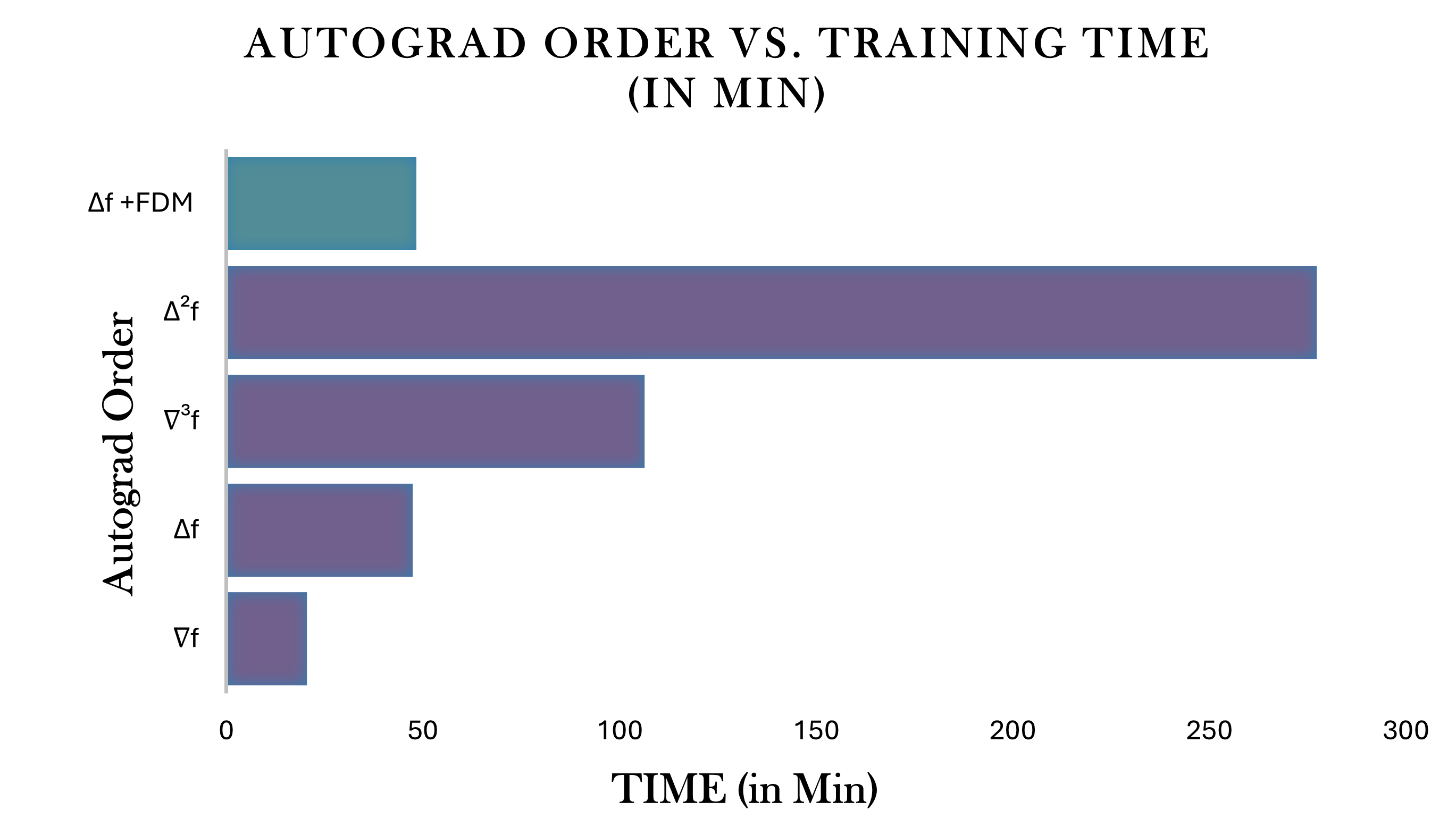}
    \caption{\textbf{Time complexity of autograd} The graph presents the time taken for 40000 steps of training with the increase in autograd order, on an RTX 3050 GPU. }
    \label{fig:order_time}
\end{figure}

The Hessian energy term requires computing fourth-order derivatives (Eq. \eqref{eq:hessian_loss}). PyTorch’s \texttt{autograd} provides a convenient and memory-efficient mechanism for automatic differentiation by constructing a dynamic computation graph, enabling backpropagation and gradient-based optimization. This however incurs significant memory and computational overhead. The peak memory consumption scales non-linearly with batch size and resolution, with fourth-order derivative computation becoming prohibitvely costly. We analyze the trade-offs between accuracy and efficiency by comparing runtime and GPU memory usage with and without the Hessian energy term. \cref{fig:hessian_opt} showcases the effect of Hessian energy optimization. To efficiently balance accuracy and computational overheads, we combine autograd with finite difference approximation, leveraging autograd to compute the second order derivative and using finite difference approximations to compute the fourth-order derivative. As shown in \cref{fig:comp_overhead} and \cref{fig:order_time}, this balance allows for feasible training times and lower computational demand.

\subsection{Ablations}
To underscore the importance of each component within our framework, we perform ablation studies, as presented in \cref{tab:ablations}. The most optimal results are seen in the highlighted row. There is a slight increase in completeness when least-squares optimization is applied. This is due to the volume reduction caused by vertex relocation. 
\begin{table}[th!]

\caption{Quantifying the impact of each component of the architecture on \textit{NewerCollege}. }
\begin{tabular}{llll}

\toprule
\textbf{SETUP} & \textbf{ACC. $\downarrow$} & \textbf{COMP.$\downarrow$} & \textbf{C-L1 $\downarrow$} \\ \midrule
MLP Network & 7.48 & 10.95 & 0.092 \\
Multi-Scale MLP & 7.01 & 10.95 & 0.091 \\
Multi-Scale MLP + Hess. Loss & 6.84 & 10.80 & 0.088 \\
\rowcolor{softgreen} 
Multi-Scale MLP + Hess. Loss + LSO & 6.52 & 10.84 & 0.087 \\ \bottomrule
\end{tabular}
\label{tab:ablations}
\end{table}

\section{Discussion}

\subsection{Numerical stability of higher-order derivatives}
\label{sec:autograd_tradeoff}

Numerical stability is a key challenge when incorporating second- and fourth-order derivatives into the loss function. The magnitude of these higher-order derivatives varies significantly compared to first-order terms, requiring careful scaling to ensure stable optimization. For second-order derivatives, we require a scaling factor of $1e^{-8}$ to maintain numerical balance with other loss terms. For fourth-order derivatives, a more aggressive scaling factor of $1e^{-11}$ is applied; without this adjustment, the excessively large gradients introduce instability, leading to mesh collapse as the optimization process rapidly distorts the underlying geometry. A potential direction for future work is to explore log transformations or adaptive gradient clipping to regulate the gradient magnitude dynamically.

\subsection{Global effects of least squares optimization}
\begin{figure}[th!]
    \centering
    \includegraphics[width=8.5cm]{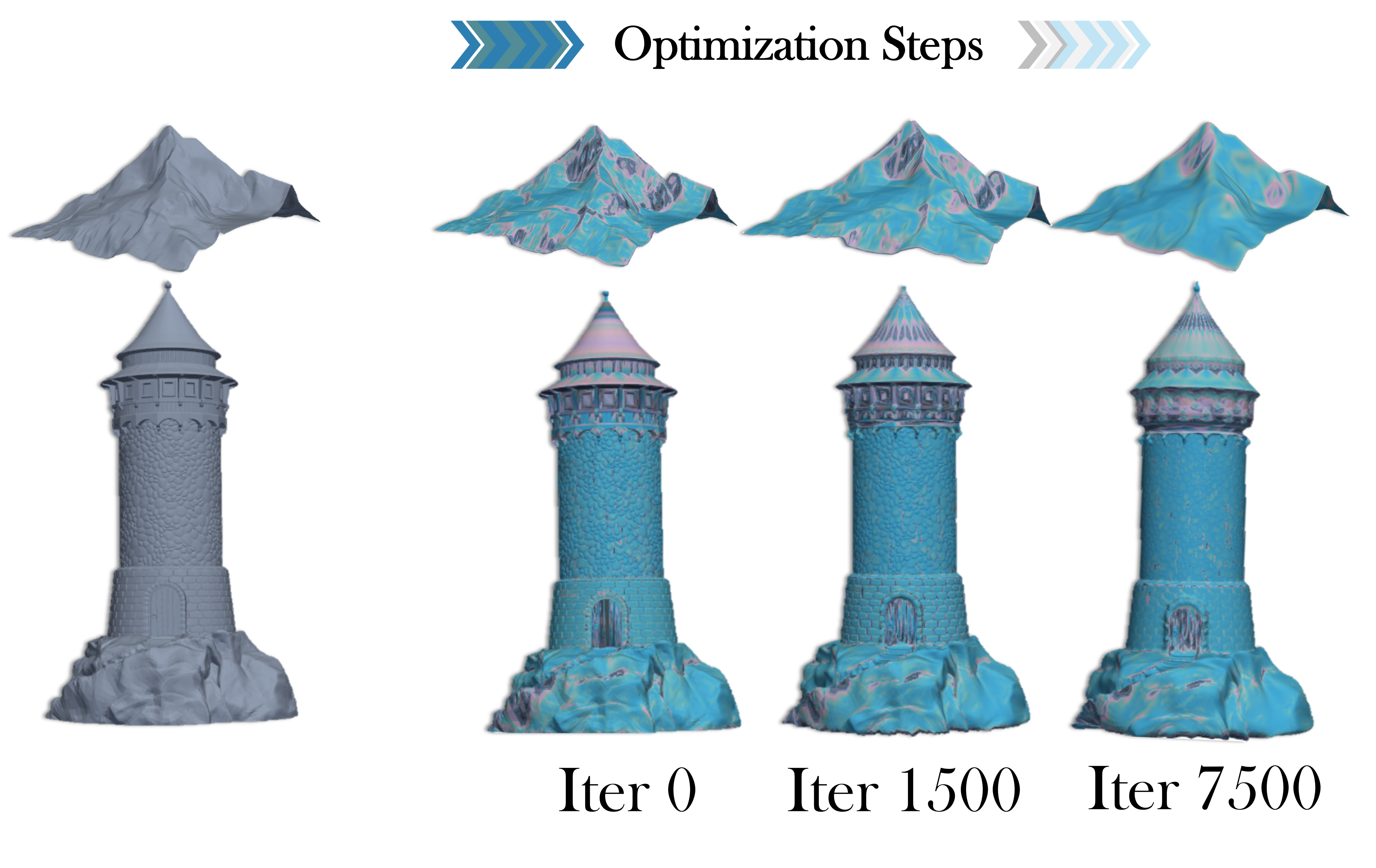}
    \caption{\textbf{Oversmoothing effects:} This figure highlights oversmoothing, observed due to repeated application of Laplace optimization. It can be observed that key features are lost due to oversmoothing.}
    \label{fig:oversmoothing}
\end{figure}

Least squares optimization is applied across the generated mesh to enforce smoothness as seen in \cref{fig:laplace-smoothing}. The number of iteration steps is treated as a hyperparameter, with optimization terminating when the F1 score stops improving. However, different regions exhibit varying curvature levels: While high-curvature areas require additional iterations for sufficient refinement, increasing the iteration count indiscriminately leads to oversmoothing in regions where sharp edges are an intentional feature rather than an artifact. This can be seen in \cref{fig:oversmoothing}, with lighter shades of blue indicating smoother regions.  Additionally, smooth regions with inherently low Laplacian values experience slow but cumulative smoothing effects, resulting in excessive surface diffusion over prolonged optimization. A more structured approach to mesh regularization could involve targeted smoothing strategies, such as semantic segmentation to classify and treat different surface regions independently or applying hard thresholds to selectively constrain smoothing in feature-preserving regions while still refining high-curvature areas.

\section{CONCLUSION}
We present a physics-informed approach to surface mapping from dense LiDAR point-cloud datasets. We show that our approach achieves greater accuracy and smoothness compared to state-of-the-art models. We present an efficient formulation of physics-based losses, optimized for speed and computational efficiency, and a test-time refinement strategy for fine-grained improvements in smoothness.

\section*{ACKNOWLEDGMENTS}
This material is based upon work supported in part by the DEVCOM Army Research Laboratory under cooperative agreement W911NF2020221.


\bibliography{root}
\bibliographystyle{ieeetr}

\end{document}